\documentclass[pra,twocolumn,showpacs,amsmath,amssymb,superscriptaddress]{revtex4}

\usepackage{graphicx}
\usepackage{dcolumn}
\usepackage{bm}

\begin{document}

\title{
High-fidelity cluster state generation for ultracold atoms in an optical lattice
}

\affiliation{
NTT Basic Research Laboratories, NTT Corporation, Atsugi 243-0198, Japan
}
\affiliation{
NTT Secure Platform Laboratories, NTT Corporation,
Musashino 180-8585, Japan
}
\affiliation{
JST, CREST, Chiyoda-ku, Tokyo 102-0075, Japan
}

\author{Kensuke Inaba}
\affiliation{
NTT Basic Research Laboratories, NTT Corporation, Atsugi 243-0198, Japan
}
\affiliation{
JST, CREST, Chiyoda-ku, Tokyo 102-0075, Japan
}

\author{Yuuki Tokunaga}
\affiliation{
NTT Secure Platform Laboratories, NTT Corporation,
Musashino 180-8585, Japan
}
\affiliation{
JST, CREST, Chiyoda-ku, Tokyo 102-0075, Japan
}
\author{Kiyoshi Tamaki}
\affiliation{
NTT Basic Research Laboratories, NTT Corporation, Atsugi 243-0198, Japan
}
\affiliation{
JST, CREST, Chiyoda-ku, Tokyo 102-0075, Japan
}

\author{Kazuhiro Igeta}
\affiliation{
NTT Basic Research Laboratories, NTT Corporation, Atsugi 243-0198, Japan
}
\affiliation{
JST, CREST, Chiyoda-ku, Tokyo 102-0075, Japan
}

\author{Makoto Yamashita}
\affiliation{
NTT Basic Research Laboratories, NTT Corporation, Atsugi 243-0198, Japan
}
\affiliation{
JST, CREST, Chiyoda-ku, Tokyo 102-0075, Japan
}

\date{\today}

\begin{abstract}
We propose a method for generating high-fidelity multipartite spin-entanglement of ultracold atoms in an optical lattice in a short operation time with a scalable manner, which is suitable for measurement-based quantum computation.
To perform the desired operations based on the perturbative spin-spin interactions, we propose to actively utilize the extra degrees of freedom (DOFs) usually neglected in the perturbative treatment but included in the Hubbard Hamiltonian of atoms, such as, (pseudo-)charge and orbital DOFs.
Our method simultaneously achieves high fidelity, short operation time, and scalability by overcoming the following fundamental problem: enhancing the interaction strength for shortening operation time breaks the perturbative condition of the interaction and inevitably induces unwanted correlations among the spin and extra DOFs.
\end{abstract}


\preprint{APS/123-QED}

\maketitle

Measurement-based quantum computation (MBQC) \cite{Briegel2009,Raussendorf2001} is a prominent method for scalable quantum information processing.
The essential ingredients of MBQC are single-qubit measurements and generation of a multipartite entangled cluster state.
The fault-tolerant MBQC requires a scalable entangling operation with fidelity beyond a threshold of $\sim$99\% within short operation time compared to coherence time \cite{Raussendorf2007,Raussendorf2007a}.
Ultracold atoms with pseudospin ({\it e.g.,} hyperfine) states in an optical lattice are promising candidates with which to implement scalable MBQC, because the large number of atoms in a lattice can sustain quantum coherence for a long time \cite{Bloch2008a}.
Atom microscope techniques are making rapid progress in the single-atom measurement \cite{Bakr2010,Weitenberg2011}.
Various entanglement generation methods with ultracold atoms have been proposed \cite{Mandel2003a,Mandel2003,Trotzky2008,Trotzky2010,Anderlini2007CEI, Vaucher2008,Duan2003,Jiang2009,YuAo2011,Simon2011QSO}. However, they have encountered intrinsic errors leading to a certain amount of a decrease in fidelity, which are serious for MBQC.
Such an error originates from the difficulty of performing the quantum operations as mentioned below.

A general method for creating a cluster state $|{\rm CS}\rangle$ is to utilize the time-evolution given by the Ising Hamiltonian $\hat{\cal H}_{\rm I}=J_{\rm I}\sum_i\hat\sigma_i^z\hat\sigma_{i+1}^z$ \cite{Raussendorf2001}:
 $|{\rm CS}\rangle=e^{ -i\tau_I\hat{\cal H}_{\rm I}/\hbar}\prod_i |+\rangle_i$, where $|+\rangle_i=(|0\rangle_i+|1\rangle_i)/\sqrt{2}$ and $\tau_I$ is the specific time $\pi\hbar/4J_{\rm I}$.
On the other hand, atoms in an optical lattice are described by the Hubbard Hamiltonian $\hat{\cal H}_{\rm Hub}$ \cite{Bloch2008a}.
The high controllability of cold atoms helps us to flexibly design $\hat{\cal H}_{\rm Hub}$ as follows: $e^{-i\tau \hat{\cal H}_{\rm Hub}/\hbar} = e^{-i\tau \hat{\cal H}_{I}/\hbar}\hat{\cal U}_{\rm ext}$, where $\hat{\cal U}_{\rm ext}$ is an extra operation.
An ideal goal is to set $\hat{\cal U}_{\rm ext}$ to be the identity operator $\hat{1}$, while a difference in the degree of freedom (DOF) between $\hat{\cal H}_I$ and $\hat{\cal H}_{\rm Hub}$ could induce $\hat{\cal U}_{\rm ext}\not=\hat{1}$.
Namely, $\hat{\cal H}_{\rm Hub}$ includes spin, pseudo-charge, and orbital DOFs, while $\hat{\cal H}_{I}$ has only a spin (qubit) DOF.

One strategy for creating spin-spin interactions is a modulation of Wannier orbital to make overlap between orbitals at neighboring sites \cite{Mandel2003a,Mandel2003,Trotzky2008,Trotzky2010,Anderlini2007CEI}.
In these schemes, the large change in orbitals of {\it qubit itself} inevitably causes errors resulting from the extra DOF.
Other strategy for avoiding the change in qubit itself builds on a perturbative interaction analogous to the Heisenberg Hamiltonian $\hat{\cal H}_{\rm Hei} (=\hat{\cal H}_{I}+\hat{\cal H}_{\rm ex})$, where $\hat{\cal H}_{\rm ex}=J_{\rm ex}\sum_i(\hat{\sigma}^x_i\hat{\sigma}^x_{i+1}+\hat{\sigma}^y_i\hat{\sigma}^y_{i+1})$.
However, the spin-exchange Hamiltonian $\hat{\cal H}_{\rm ex}$ will be a source of $\hat{\cal U}_{\rm ext}$ even though it can be described by the spin DOF.
Another (worse) problem is that the perturbative nature results in very weak interaction strengths and requires a long operation time.
One approach to the two problems is enhancing $J_I$ for achieving $J_I\gg J_{\rm ex}$ \cite{Duan2003,Jiang2009,YuAo2011,Simon2011QSO}.
However, it
causes the breakdown of the perturbative condition;
namely,  when achieving a short operation time, the effects of extra DOFs neglected in the perturbative treatment results in $\hat{\cal U}_{\rm ext}\not=\hat{1}$ that could decrease fidelity.
This tradeoff between short operation time and high fidelity is a fundamental problem for the entanglement generations.
Moreover, serious difficulty emerges when the number of qubits becomes large due to the $\hat{\cal U}_{\rm ext}$ caused by many-body correlations and collective effects, which degrade scalability of the operation.

In this paper, we propose that an active control of the extra DOFs without neglecting them provides a solution of the above problems.
We find that, even though $\hat{\cal U}_{\rm ext}\not=\hat{1}$, when $\hat{\cal U}_{\rm ext}$ is designed to be controllable, fidelity reaches very high value of $\gtrsim 0.99$ in a short time.
Here, we choose a spin DOF of fermionic atoms in the lowest orbital as qubit, and we utilize the extra Hilbert space spanned by the higher orbitals as an ancillary subspace.
The extra subspace allows us to naturally create an Ising interaction with a tunable $J_I$. The distinguishability of the two spaces assures the disappearance of the $\hat{\cal U}_{\rm ext}$ originating from spin-exchange terms.
Using {\it ab-initio} numerical simulations, however, we find that another type of $\hat{\cal U}_{\rm ext}(\not=\hat{1})$ is introduced inevitably whenever $J_I$ is enhanced to achieve a short operation time.
We also clarify that this $\hat{\cal U}_{\rm ext}$ induces an extra excitation to the ancillary subspace, which originates from the Rabi-oscillation-like mechanism, but it does not matter to fidelity because it can be controlled as follows.
Thanks to the absence of the spin exchange terms, we can minimize a decrease in fidelity without the time-fidelity tradeoff by controlling the oscillatory dynamics resulting from this $\hat{\cal U}_{\rm ext}$.
We further improve the fidelity by detecting the states in the ancillary space combined with  post selection.
Moreover, for scalability, we propose a pairwise entanglement generation scheme for keeping the properties of $\hat{\cal U}_{\rm ext}$ unchanged with increasing number of qubits.
Our basic concept is to control both the intended operation $e^{-i\tau \hat{\cal H}_{I}/\hbar}$ and the unavoidable extra operation $\hat{\cal U}_{\rm ext}$.
This general concept can be applied to the other systems if we have a controllable DOF that is suitable for the ancillary space, such as Wannier orbitals.

The Hubbard Hamiltonian discussed  here is written as $\hat{\cal H}_{\rm Hub}=\sum_{\langle ij\rangle}\sum_{\sigma\alpha}J_{\alpha\alpha} \hat{c}^\dag_{i\sigma\alpha} \hat{c}_{{j}\sigma\alpha}+\sum_i\sum_{\sigma\alpha} \varepsilon_\alpha \hat{c}^\dag_{i\sigma\alpha}\hat{c}_{i\sigma\alpha}+\sum_{i}\sum_{\alpha\beta} U_{\alpha\beta} ({\hat{c}}^\dag_{i\uparrow\alpha}{\hat{c}}_{i\uparrow\alpha}{\hat{c}}^\dag_{i\downarrow\beta}{\hat{c}}_{i\downarrow\beta}+{\hat{c}}^\dag_{i\uparrow\alpha}{\hat{c}}_{i\uparrow\beta}{\hat{c}}^\dag_{i\downarrow\beta}{\hat{c}}_{i\downarrow\alpha})+H.c.,$ where ${\hat{c}}^\dag_{i\sigma\alpha} ({\hat{c}}_{i\sigma\alpha})$ are the creation (annihilation) operator of a fermion with a spin $\sigma(=\uparrow, \downarrow)$ and an orbital $\alpha(=1,2,\cdots)$ at the $i$th site.
The parameters $J_{\alpha\alpha}$, $\varepsilon_\alpha$, and $U_{\alpha\beta}$ can be derived from the optical lattice potential:
$V_0 [\cos(x/a)+\cos(y/a)+\cos(z/a)]$, where $V_0$ is the lattice depth, and $a$ is the lattice distance.
Note that $U_{\alpha\beta}$ is determined by considering the s-wave scattering length of the atoms $a_{S}$. (see Appendix \ref{appA})
In what follows, we choose a large $V_0$ to set $J_{\alpha\beta}\ll U_{\alpha\beta}$ leading to the localization of atoms and then define localized single atoms with $\sigma=\uparrow\,$($\downarrow$) occupying the lowest orbital $\alpha=1$ at the $i$th site as the qubits, and  we express its state by $|0\rangle_i\,$($|1\rangle_i$).

\begin{figure}[tb]
\includegraphics[width=8.5cm]{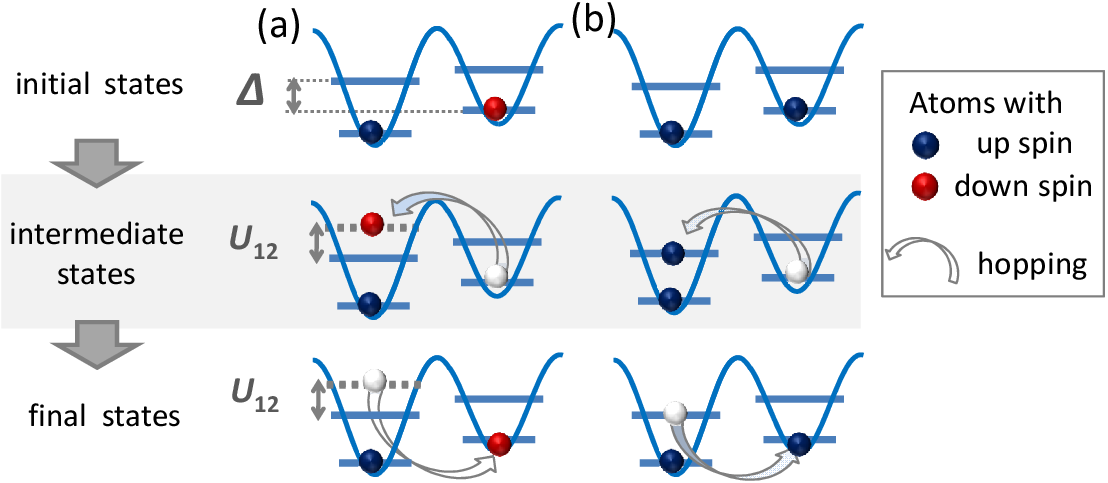}
\caption{
(a) and (b) Second order virtual hopping processes for two atoms with different and same spins, respectively.
Blue solid lines represent the energy levels of the lowest and the second lowest orbitals.
The gray dotted line shows the energy shift caused by the interaction.
}
\label{fig_bi}
\end{figure}

We first discuss the way to create spin-spin interactions based on a perturbative mechanism, such as the Heisenberg interactions $\hat{\cal H}_{\rm Hei}$ \cite{Anderson}.
The previous studies obtained the Ising interactions in the limit of $J_I\gg J_{\rm ex}$ based on $\hat{\cal H}_{\rm Hei}$ \cite{Duan2003,Jiang2009,YuAo2011,Simon2011QSO}.
In contrast, we propose to  naturally create a pure Ising interaction without spin exchange terms by utilizing the orbital DOF.
We use a particular kind of perturbation induced by the {\it interorbital} hoppings $J_{12}$.

Figure \ref{fig_bi} schematically represents the $J_{12}$-induced perturbative processes with the initial states of two localized atoms with different and same spins shown in the top panels of the column (a) and (b), respectively.
The middle panels show the virtual intermediate states.
For the atoms with different spins, the interorbital interaction $U_{12}$ causes the energy shift depicted by the gray dotted line, while there is no such  energy  shift for the atoms with same spins due to the Pauli principle.
The bottom panels are the final states, which are the same as the initial states (except for a phase difference).
We can see that the virtual transitions between distinguishable orbitals naturally yield the Ising interaction.
Importantly, there is no spin-exchange processes.
Note that the similar perturbation processes caused by the intraorbital $J_{11}$ intrinsically induce both $\hat{\cal H}_{I}$ and $\hat{\cal H}_{\rm ex}$ due to the SU(2) symmetry.

Let us explain an implementation how to induce the interorbital hopping.
We add an extra two-site period potential given by $V_0'\cos(x/2a+\pi/2)$ with the potential depth $V_0'$ to the original lattice.
It leads to the additional Hamiltonian written as $\hat{\cal H}'=\sum_{i\sigma\alpha}\varepsilon'_{i,\alpha} \hat{c}^\dag_{i\sigma\alpha}\hat{c}_{i\sigma\alpha}+\sum_{\langle ij \rangle\sigma\alpha\beta}J'_{i,\alpha\beta}\hat{c}^\dag_{i\sigma\alpha}\hat{c}_{j\sigma\beta}$.
As discussed in Appendix~\ref{appB}, the onsite potential $\varepsilon'_{i,\alpha}$ exhibits staggered modulations; and
the hopping matrices $J'_{i,\alpha\beta}$ have the following useful feature.
The interorbital $J'_{i,12}$ has a large magnitude, while the intraorbital elements, such as $J'_{i,11}$, are negligibly small, which is key to the suppression of $\hat{\cal H}_{\rm ex}$.
The additional potential results in the Ising interaction with $J_{\rm I}\propto \frac{|J'_{12}|^2}{\Delta+U_{12}}-\frac{|J'_{12}|^2}{\Delta}$ via the mechanism mentioned above, where $\Delta=\varepsilon_2-\varepsilon_1+\varepsilon'_{i,2}-\varepsilon'_{i+1,1}$.

The interaction strength can be increased if the resonant condition $\Delta+U_{12} \sim 0$ is satisfied, which can be
achieved  by controlling $U_{12}(\propto a_{S})$ via a Feshbach resonance \cite{Regal2003} and also  by tuning $\varepsilon'_{i,\alpha}$ via the change in $V_0'$.
The resonance allows us to reduce the time required for preparing cluster states.
However, this corresponds to the breakdown of the perturbative assumption used in the above discussions.
By calculating the actual dynamics of atoms without any perturbative approximations, we clarify that the breakdown of perturbation introduces an extra operation $\hat{\cal U}_{\rm ext}$, and we also reveal that this $\hat{\cal U}_{\rm ext}$ can be handled thanks to the absence of $\hat{\cal H}_{\rm ex}$.

\begin{figure}[tb]
\includegraphics[width=8.cm]{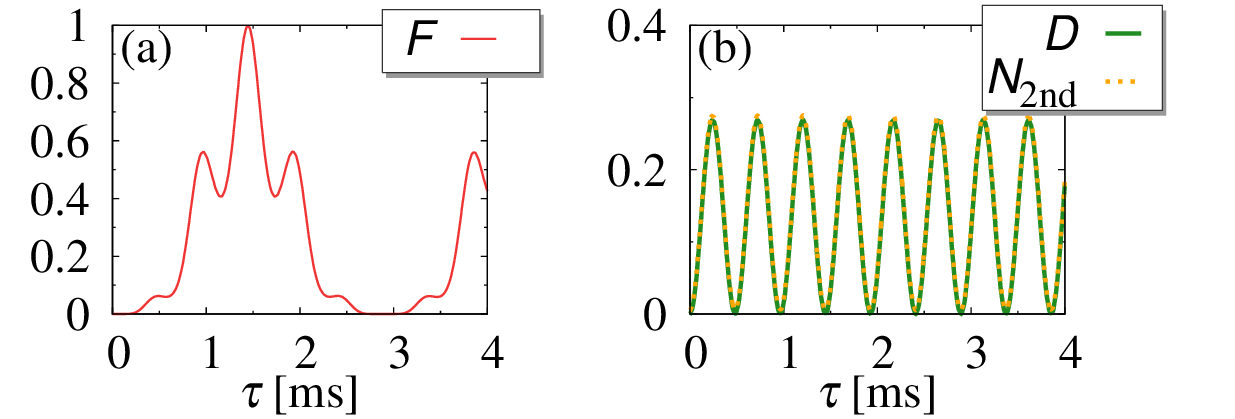}
\caption{Simulation results for the 2-qubit system with $V_0=10E_r$ and $V_0'=6.2E_r$. (a) Fidelity $F$ as a function of time $\tau$. (b) Expectation values of the number of doubly occupied sites  $D$ and the atoms with the second lowest orbital $N_{\rm 2nd}$. The two curves overlap with each other.
}
\label{fig_fid}
\end{figure}

We numerically simulate the time evolution of atoms written as $|\phi(\tau)\rangle=\exp\{i(\tau/\hbar)(\hat{\cal H}_{\rm Hub}+\hat{\cal H}')\}\prod_i|+\rangle_i$ using an exact diagonalization method.
First of all, we investigate a 2-qubit (2-atom) system of $^{40}$K atoms with the following realistic parameters; $a=413$ nm, $a_{S}=-50$ nm, $V_0=10 E_r$ and $V_0'=6.2 E_r$, where $E_r$ is the recoil energy.
Two hyperfine states, $|F,m_F\rangle=|9/2,-9/2\rangle$ and $|9/2,-7/2\rangle$, are considered as a spin DOF.
We confirm that the third lowest and the higher orbitals are negligible because their energy levels are far off-resonant from that of the lowest orbital.
In Fig.\,\ref{fig_fid}\,(a), we show the calculated fidelity $F\equiv|\langle{\rm CS}|\phi(\tau)\rangle|^2$ as a function of time $\tau$ \cite{NoteBell}.
We find that the $F$ curve is characterized by two types of oscillations with periods of about 3 and 0.5 ms.
The long period oscillation is obviously induced by the Ising operation $e^{-i\tau\hat{\cal H}_I/\hbar}$.
The short one suggests the existence of the extra operation $\hat{\cal U}_{\rm ext}(\not=\hat{1})$.

To reveal the origin of $\hat{\cal U}_{\rm ext}$, we calculate expectation values of the number of doubly occupied sites $D$ and the number of atoms in the second lowest orbital $N_{\rm 2nd}$ as a function of $\tau$.
As shown in Fig.\,\ref{fig_fid}\,(b), two quantities $D$ and $N_{\rm 2nd}$ agree exactly, and the oscillation period of them is 0.5\,ms.
These results mean that the $\hat{\cal U}_{\rm ext}$ inducing the short period oscillation is characterized by an actual excitation of the intermediate state shown in the middle panel of Fig.\,\ref{fig_bi}\,(a).
This resonance dynamics can be understood as the Rabi-oscillation-like phenomenon, and thus, we can determine the time period from the inverse of the effective Rabi frequency given by $\Omega_R=\sqrt{\delta^2+4|J_{12}'|^{2}}/\hbar$ with $\delta\equiv \Delta+U_{12}$ corresponding to the detuning from the resonant condition (see Appendix~\ref{appC}).
Note that the intermediate state is assumed to be virtual in the simple perturbation theory.
Such a dynamics beyond the perturbative assumption is induced in return for enhancing $J_I$,
which is the tradeoff between the high fidelity and short operation time resulting from the breakdown of the perturbation.

\begin{figure}[tb]
\includegraphics[width=9cm]{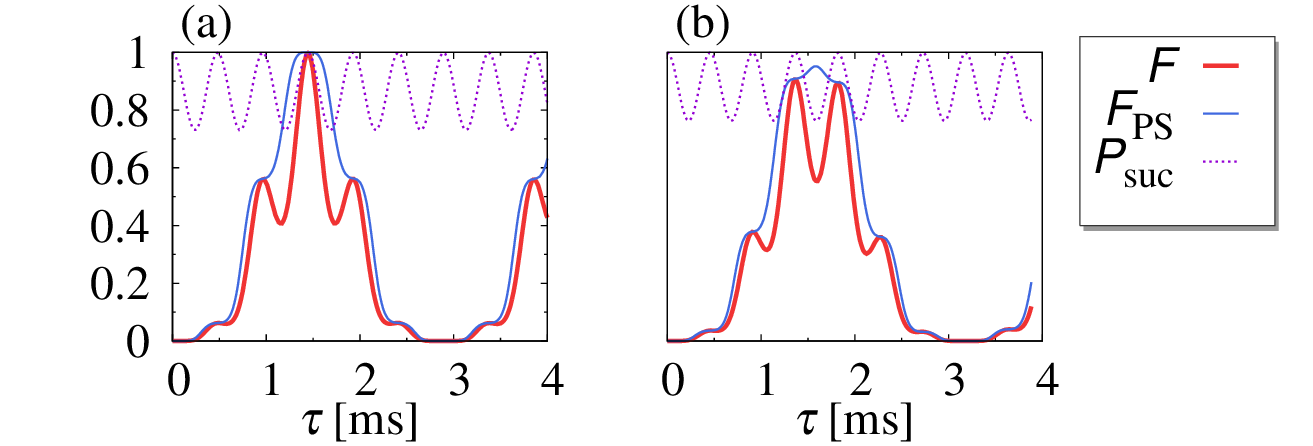}
\caption{Fidelity $F$ and that with the post selection $F_{\rm PS}$, and success probability $P_{\rm suc}$ of the 2-qubit system with $V_0=10 E_r$ for (a) $V_0'=6.2 E_r$ and (b) $6.23 E_r$.
}
\label{fig_fidPS}
\end{figure}

We provide that this fundamental problem can be overcome by controlling both the two oscillations.
To explain this point, we show two calculated $F$ curves with the same parameters except for $V_0'=6.2$ and $6.23 E_r$ shown as red thick lines in Fig.~\ref{fig_fidPS}.
By comparing two $F$ curves, we find that the short and the long period oscillations are {\it inphase} for $V_0'=6.2 E_r$, while {\it opposite inphase} for $V_0'=6.23 E_r$.
For the former case, we achieve very high fidelity $F=0.997$ in a short time of $1.5$~ms; while for the latter case, $F$ reaches up to 0.907 at the highest at around $\tau=1.4$~ms.
These results reveal that a decrease in fidelity in return for shortening the operation time can be minimized by setting two oscillations caused by $\hat{\cal U}_{\rm ext}$ and $e^{-i\tau \hat{\cal H}_I/\hbar}$ inphase, and this can be done by tuning the parameter $V_0'$.

This phase-tuning scheme allows us to resolve the difficulty resulting from the breakdown of the perturbation.
Instead, the phase-tuning condition $ \ell J_I=(m-1/2)\hbar\Omega_R$ with integers $\ell$ and $m$ of $\ge 1$ is imposed for simultaneously achieving short operation time and high fidelity.
The above simulation used $\ell=3$ and $m=1$, while the shortest $\tau_I$ can be obtained by $\ell=m=1$ (see Appendix~\ref{appD}).
Importantly, the phase-tuning scheme can be implemented because the present $\hat{\cal U}_{\rm ext}$ induces only one type of the excitations clarified above.
For instance, when $\hat{\cal U}_{\rm ext}$ further includes $\hat{\cal H}_{\rm ex}$, three kinds of oscillations will appear in the $F$ curves, causing a great difficulty of the phase tuning.

To remove tiny errors still remaining after applying the above scheme, we propose the following postselection scheme:
first, detect $D$ by using, {\it e.g.}, photo-association spectroscopy \cite{Rom2004,Sugawa2011}, or measure $N_{\rm 2nd}$ with, {\it e.g.},  orbital blockade spectroscopy \cite{Bakr2011};
after that, determine if the operation is a success (failure) by the measurement outcome suggesting the absence (existence) of the intermediate states.
Figure\,\ref{fig_fidPS} shows fidelity with the post selection $F_{\rm PS}$ (blue thin line) and the success probability $P_{\rm suc}$ (purple dashed line).
By comparing $F_{\rm PS}$ with $F$, we find that the postselection scheme improves fidelity.
Another interesting feature is that $F_{\rm PS}$ curves are more smooth than $F$ curves, and thus, the time of being high fidelity is lengthened (see Appendix~\ref{appE}).
For $V_0=6.2 E_r$, an extremely high fidelity of $F_{\rm PS}\sim1$ is achieved in return for the failure probability of $1-P_{\rm suc}\sim0.002$, suggesting that almost all of the remaining error is detectable in the ancillary subspace.
In contrast, for $V_0=6.23 E_r$, we obtain $F_{\rm PS}=$0.95 with $1-P_{\rm suc}=0.211$.
The post selection cannot work effectively without the phase tuning, which is due to the uncommutativity between $e^{-i\tau \hat{\cal H}_I}$ and $\hat{\cal U}_{\rm ext}$.
These results also indicate the importance of designing $\hat{\cal U}_{\rm ext}$ to include only one kind of excitations.

\begin{figure}[tb]
\includegraphics[width=8.5cm]{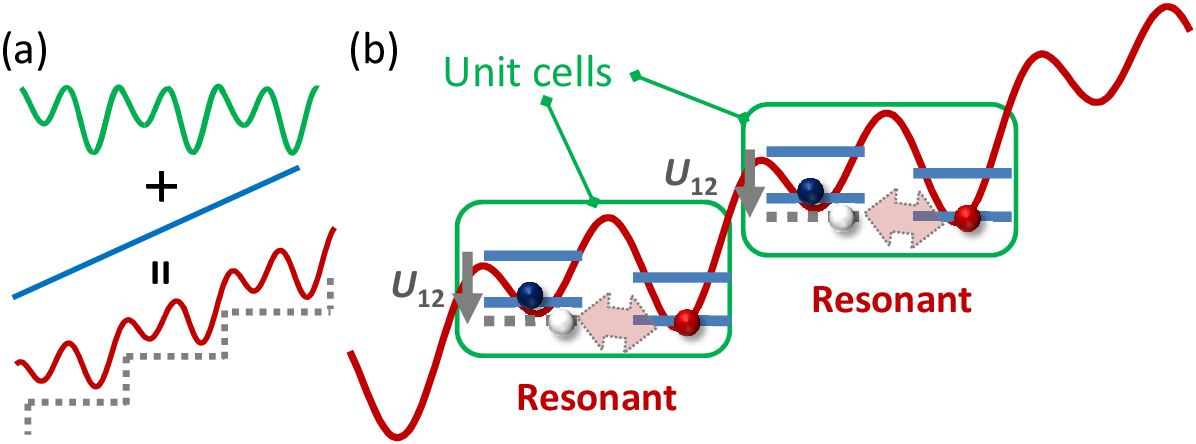}
\caption{(a) Overall diagram of a linear potential modulation in addition to the lattice potential.
The dotted line provides a stepwise structure to guide the eyes.
(b) A schematic view of the pairwise scheme, where the Ising interaction is resonantly enhanced only inside the unit cells.
}
\label{fig_scale}
\end{figure}
\begin{figure*}%[tb]
\includegraphics[width=10cm]{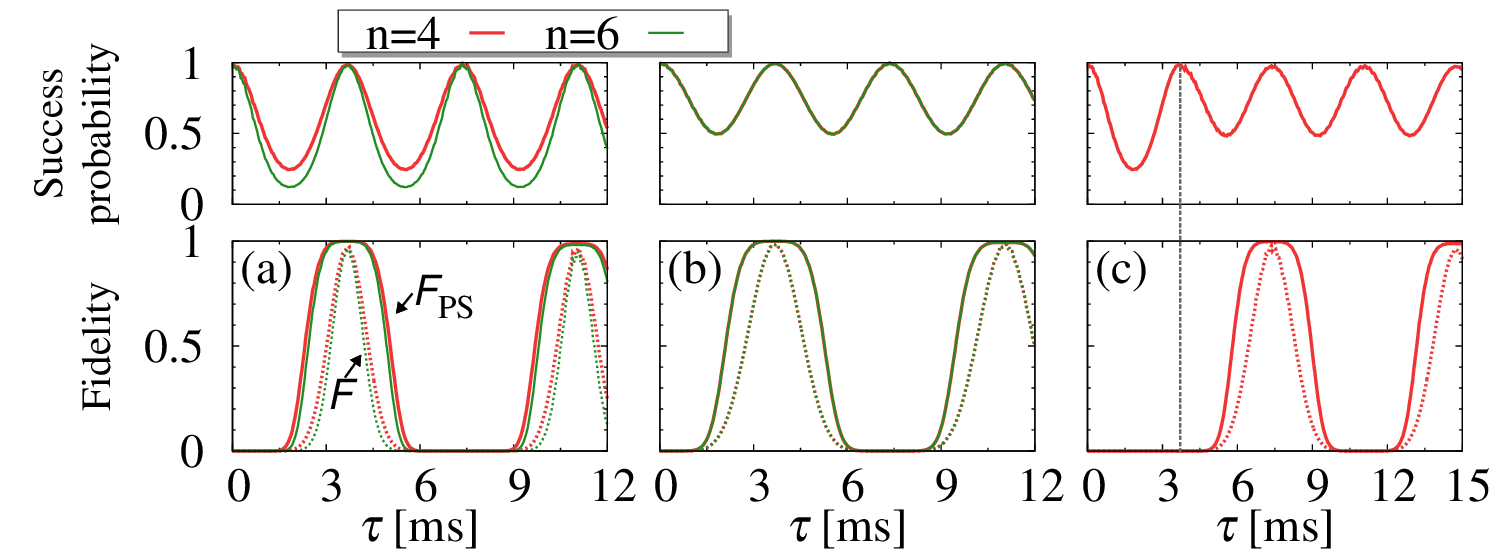}
\caption{
Simulation results using the pairwise scheme for $4$- and 6-qubit systems with $V_0=18E_r$ and $V_0'=4E_r$.
(a) Fidelity $F$ and $F_{\rm PS}$, and success probability $P_{\rm suc}$.
(b) Rescaled quantities  $F_{}^{2/n}$, $F_{\rm PS}^{2/n}$, and $P_{\rm suc}^{2/n}$, where the curves for different $n=4$ and $6$ overlap each other.
(c) Same quantities with the unit cell shift for $n=4$.
The gray dashed line represents the time $\tau=3.7$ ms at which we move the unit cells.
}
\label{fig_mdep}
\end{figure*}
Next, let us discuss scalability of our schemes by simulating $n$-qubit systems for $n>2$.
When we apply our scheme to 4-qubit system in the same way as the above, we obtain the fidelity of $F_{\rm PS}\sim 0.71$ with $P_{\rm suc}\sim 0.931$.
This crucial decrease in the fidelity is attributed to another kind of $\hat{\cal U}_{\rm ext}$ not included in the 2-qubit simulations.
For instance, in the intermediate state shown in Fig.\,\ref{fig_bi}\,(a), the excited atom does not always go back to the initial site, but instead further moves to the third lattice site.
This means that, due to the collective motion of atoms, $\hat{\cal U}_{\rm ext}$ changes drastically as $n$ increases; and thus, $\hat{\cal U}_{\rm ext}$ cannot be controlled at all by the above two schemes for a large $n$.
Note that the collective $\hat{\cal U}_{\rm ext}$ could be resonantly enhanced, meaning that the perturbative nature raises a fundamental difficulty of simultaneous achievement of high fidelity, short operation time, and scalability.

To solve this problem, we further propose the pairwise scheme: divide the system into a set of effective 2-qubit systems; and then independently generate an entanglement in each 2-qubit system, where $\hat{\cal U}_{\rm ext}$ is expected to be the same as those in the 2-qubit simulation mentioned above.
This scheme is implemented by adding a linear potential $gx$ with a gradient $g$.
Figure \ref{fig_scale}\,(a) schematically represents the linear potential modulation.
Here we find a characteristic stepwise structure.
As shown in Fig.\,\ref{fig_scale}\,(b), we define each step consisting of two lattice sites as a unit cell, and
 properly tune the resonant condition $\delta_{}\sim0$ satisfied only inside the unit cells.
Consequently,
the cluster state in each unit cell can be generated simultaneously and separately within a short operation time.
It does not matter if the commutative Ising operation $e^{-i\tau\hat{\cal H}_I}$ is pairwise or not.
In contrast, we should make the uncommutative $\hat{\cal U}_{\rm ext}$ pairwise, and
the scalability of entanglement generations considering $\hat{\cal U}_{\rm ext}$ should be carefully examined by numerical simulations.

In Fig.\,\ref{fig_mdep}\,(a), we present the fidelity $F$ between  $|\phi(\tau)\rangle$ and a product state of the cluster state in unit cells for $n=4$ and $6$ with $V_0=18 E_r$ and $V_0'=4 E_r$ \cite{NoteGradient}.
The phase-tuning condition is satisfied with $\ell=m=1$.
High-fidelity entangled states for $n=4$\,($6$) are generated with or without the post selection as $F_{\rm PS}\sim0.999\,(0.998)$ with $P_{\rm suc}\sim0.987\,$($0.980$) or $F\sim0.973$\,($0.960$), respectively.
Figure\,\ref{fig_mdep}\,(b) shows that rescaled quantities $F^{2/n}$, $F_{\rm PS}^{2/n}$ and $P_{\rm suc}^{2/n}$.
For all of these quantities, the rescaled curves for $n=4$ and $6$ overlap each other, suggesting that $\hat{\cal U}_{\rm ext}$ is almost perfectly pairwise.
We obtain the unit-cell fidelity of $\tilde{F}_{\rm PS} \sim 0.999$ with $\tilde{P}_{\rm suc} \sim 0.993$ or $\tilde{F}_{}\sim 0.987$.
A combination of the three schemes achieves a fidelity beyond $\sim$99\% with scalability.

We further extend our scheme for generating multipartite entangled states over whole 1D lattice sites.
After the set of 2-qubit cluster states are prepared as mentioned above, we move unit cells toward one site right or left by changing the sign $g$ or the relative phase $\theta=\pi/2 \to 3\pi/2$, and then again perform pairwise entanglement generation.
Figure\,\ref{fig_mdep}\,(c) shows simulation results for $n=4$ with the same parameters as  above, where unit cells are shifted at time $\tau=3.7$ ms indicated by the gray dashed line.
We find only the same kind of $\hat{\cal U}_{\rm ext}$ as the above even after $\tau=3.7$\,ms, and thus we finally obtain very high-fidelity cluster states:
$F_{\rm PS}=0.998 (\sim \tilde{F}_{\rm PS}^3)$ with $P_{\rm suc}=0.979 (\sim \tilde{P}_{\rm suc}^3)$ or $F\sim 0.961(\sim \tilde{F}_{}^3)$.

In the similar way to the above, we can create 2D and 3D cluster states with the successive 1D entanglement generation by changing the direction, which will be useful for fault tolerant MBQC \cite{Raussendorf2007a,Raussendorf2007}.
Each 1D operation is performed by inducing a large $J_{12}'$ along a certain direction only, which
 suppresses the $\hat{\cal U}_{\rm ext}$ causing the unwanted excitation along the other directions.
Our schemes are also suitable for the loss tolerant MBQC scheme \cite{Stace2009,Barrett2010}.
The failure events of the measurement used in the postselection scheme can be regarded as losses of qubits.
It is useful to enhance the fidelity in return for the increasing losses, because MBQC is more robust against losses than errors \cite{Stace2009,Barrett2010}.

In summary, we propose a method for generating spin-entanglement of atoms in an optical lattice by controlling the Wannier orbital.
Our method allows us to overcome the fundamental limit of the operations created by the perturbative interactions and achieves three properties of entanglement generations required for MBQC: high fidelity, scalability, and short operation time.
Our basic concept is to utilize the extra (orbital and charge) DOFs neglected for the perturbative treatment to precisely control the qubit (spin) DOF.
Such a general idea can be applied to various problems.
As a prospect, we note that our tunable magnetic interaction can be employed for quantum simulations of magnetism, for instance, demonstration of the N\'eel transition.
We also remark that Wannier orbital controlling can be realized in both fermionic and bosonic systems.

\begin{acknowledgments}
We thank Y. Takahashi, K. Azuma, Y. Matsuzaki, and Y. Tokura for valuable discussions.
\end{acknowledgments}

\appendix
\setcounter{figure}{0}
\renewcommand{\thefigure}{\Alph{figure}}

\section{Derivation of the Hubbard Hamiltonian} \label{appA}
Here, we explain the derivation of the Hubbard Hamiltonian of our system, fermionic atoms with two spins in an optical lattice.
The non-interacting Hamiltonian of the atoms is given by $\hat{\cal H}_0= -\sum_i\frac{\hbar^2 \hat{\boldsymbol{\nabla}}_i^2}{2M}+V_0 [\cos(\hat{x}_i/a)+\cos(\hat{y}_i/a)+\cos(\hat{z}_i/a)]$, where $V_0$ is the lattice depth, $a$ is the lattice distance, and $M$ is the mass of the atoms.
We solve $\hat{\cal H}_0$ numerically, and calculate Bloch and Wannier orbitals. %the eigenvectors and their Fourier transformation, namely,
The  Wannier orbitals are used to derive the full Hamiltonian, $\hat{\cal H}=\hat{\cal H}_0+\hat{\cal H}_{\rm int}$, in the second quantization framework.
Here, the interaction term $\hat{\cal H}_{\rm int}$ is given by $\frac{4\pi\hbar^2a_S}{M} \sum_{i,j}\delta(\hat{\bf r}_i-\hat{\bf r}_j)$ with the scattering length $a_{S}$.
The second quantized expression of $\hat{\cal H}$ is the multiorbital Hubbard Hamiltonian written as $\hat{\cal H}=\sum_{\langle ij\rangle}\sum_{\sigma\alpha}J_{\alpha\alpha} \hat{c}^\dag_{i\sigma\alpha} \hat{c}_{{j}\sigma\alpha}+\sum_i\sum_{\sigma\alpha} \varepsilon_\alpha \hat{c}^\dag_{i\sigma\alpha}\hat{c}_{i\sigma\alpha}+\sum_{i}\sum_{\alpha\beta} U_{\alpha\beta} (\hat{c}^\dag_{i\uparrow\alpha}\hat{c}_{i\uparrow\alpha}\hat{c}^\dag_{i\downarrow\beta}\hat{c}_{i\downarrow\beta}+\hat{c}^\dag_{i\uparrow\alpha}\hat{c}_{i\uparrow\beta}\hat{c}^\dag_{i\downarrow\beta}\hat{c}_{i\downarrow\alpha})+H.c.,$
where $\hat{c}^\dag_{i\sigma\alpha} (\hat{c}_{i\sigma\alpha})$ are the creation (annihilation) operator of a fermion with a spin $\sigma(=\uparrow, \downarrow)$ and an orbital $\alpha(=1,2,\cdots)$ at the $i$th site, and the subscript $\langle ij \rangle$ means the summation over the nearest-neighbor sites.
The parameters $\varepsilon_\alpha$, $J_{\alpha\beta}$, and $U_{\alpha\beta}$ are determined by  {\it ab initio} calculations with given $M$, $V_0$, $a$ and $a_{S}$.
We note that interorbital hopping matrices vanish due to orthogonality of the set of the Bloch orbitals, {\it i.e.}, $J_{\alpha\not=\beta}=0$.
We assume that the trapping potential $V_{trap} \hat{x}^2$ is absent from our model. As  recently demonstrated in the experiment \cite{Will2010}, the trapping potential can be canceled out by superimposing the optical potentials that are generated by the red and blue detuned lasers.

%
 %, where $\sigma_i^z|0 (1)\rangle_i=\pm|0 (1)\rangle_i$.

\section{Hamiltonian of the additional potentials}\label{appB}

\begin{figure*}
\includegraphics[width=10cm]{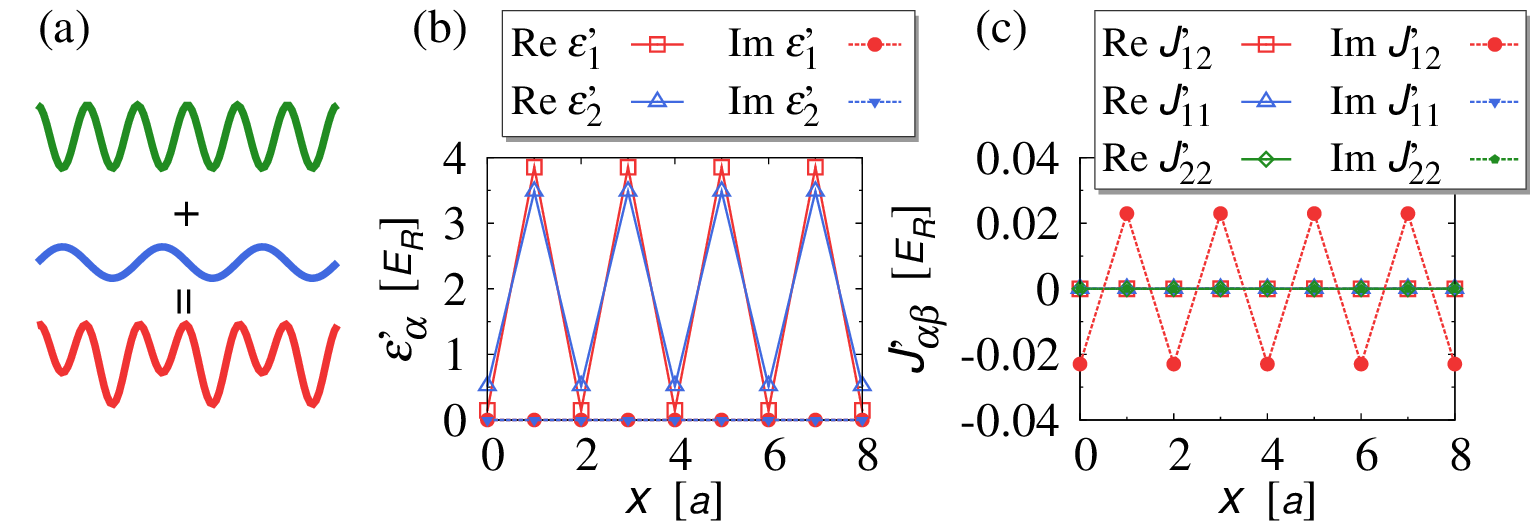}
\caption{
(a) A schematic diagram of a lattice potential modulation induced by the additional two-site period potential.
(b) and (c) The parameters of the additional potential Hamiltonian ${\cal H}'$ for $V_0=15 E_r$ and $V'_0=4 E_r$.
The open (solid) symbols represent the real (imaginary) part of these parameters.
Onsite potential terms $\varepsilon'_{i,\alpha}$ are real number, ${\rm Im}\varepsilon'_{i,\alpha}=0$.
All elements of $J_{\alpha\beta}'$ except for Im$J_{12}'$ are close to zero.
}
\label{fig_prm}\label{fig_pot}
\end{figure*}

To create an Ising interaction, we add a two-site period lattice potential, $V_0'\cos(\hat{x}/2a+\pi/2)$, to the original optical lattice.
Figure \ref{fig_pot}~(a) shows schematically how the combination of the two potentials modifies the lattice landscape.
An extra lattice potential leads to the additional Hamiltonian written as $\hat{\cal H}'=\sum_{i\sigma\alpha}\varepsilon'_{i,\alpha} \hat{c}^\dag_{i\sigma\alpha}\hat{c}_{i\sigma\alpha}+\sum_{\langle ij \rangle\sigma\alpha\beta}J'_{i,\alpha\beta}\hat{c}^\dag_{i\sigma\alpha}\hat{c}_{j\sigma\beta}$.
The two-site period potential can be rewritten by $V_0'(e^{iG/2 \hat{x}}-e^{-iG/2 \hat{x}})/2$ with the reciprocal vector $G=2\pi/a$. % with $\theta=\pi/2$. %where $V_0'$ is the po
Thus, superposing the two-site period potential causes the scattering which transfers the half-reciprocal wavevector $G/2$.
As a result, the orthogonality of the Bloch orbitals is broken, and the mixing of the Bloch orbitals is induced by the additional potential.
A numerical calculation that determines $\varepsilon'_{i,\alpha}$ and $J'_{i,\alpha\beta}$ allows us to understand the details of the orbital mixing effects.
We find that this modulation is commensurate with the original lattice so that the additional Hamiltonian $\hat{\cal H}'$ has an interesting and useful feature  as mentioned in the main text:
Namely, strong interorbital coupling $J_{12}$ is induced without enhancing the intraorbital couplings $J_{11}$ and $J_{22}$. %between orbital-$1$ and $2$

In Fig.\,\ref{fig_prm}\,(b), the onsite potential $\varepsilon'_{i,\alpha}$ exhibits staggered modulations as expected from the total potential shape shown in Fig.~\ref{fig_pot}~(a).
On the other hand, as regards the hopping matrices $J'_{i,\alpha\beta}$ shown in Fig.\,\ref{fig_prm}\,(c), we found an interesting and useful feature.
The interorbital $J'_{i,12}$ has a large magnitude, while the other elements $J'_{i,\alpha\beta}$ are negligibly small. % ($G=2\pi/a$).
Such a striking situation is realized when $\theta=\pi/2+\ell\pi$ with integer $\ell$.
As $\theta$ deviates from these values, some elements of $J'_{i,\alpha\beta}$ become non-negligible,
which causes a decrease in fidelity.

\begin{figure*}
\includegraphics[width=11cm]{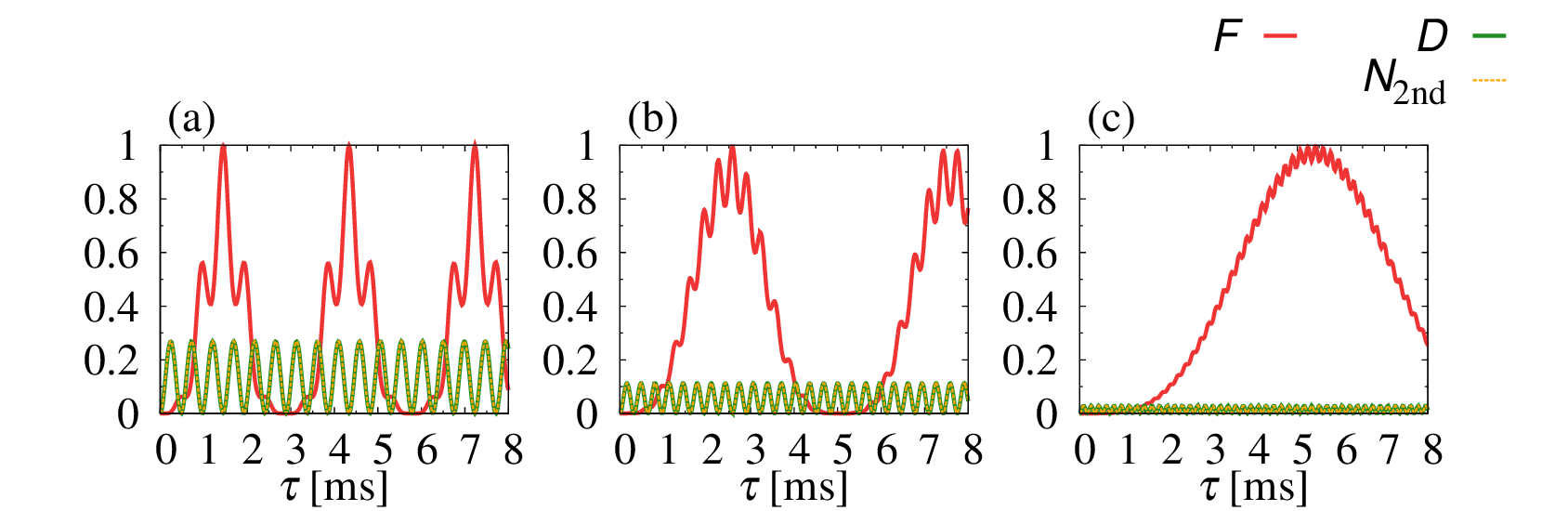}
\caption{ Simulation results for the 2-qubit system with $V_0=10\,E_r$ for $V_0'=$6.2\,$E_r$, 6.0\,$E_r$, and 5.5\,$E_r$.
}
\label{fig_fid_rabi}
\end{figure*}

\section{Resonant excitation caused by the extra operation}\label{appC}
Here, we explain in more detail the physical mechanism of the oscillatory behavior caused by the extra operation $\hat{\cal U}_{\rm ext}$.
We present results of the two-qubit simulations with the same parameters as those in Figs. 2 and 3 in the main text except for $V_0'=6.2\,E_r$, 6.0\,$E_r$, and 5.5\,$E_r$, where we use $a=413$ nm, $a_{S}=-50$ nm, and $V_0=10 E_r$.
Figure~\ref{fig_fid_rabi} shows the calculated fidelity $F$, and expectation values of the number of doubly occupied sites $D$ and the number of atoms in the second lowest orbital $N_{\rm 2nd}$.

As mentioned in the main text, for all of the results, we find two types of oscillatory behavior in $F$ curves.
The long period oscillation is caused by the Ising operation $e^{-i\tau\hat{\cal H}_I/\hbar}$, and thus the oscillation period is characterized by the Ising interaction $J_{\rm I}\propto \frac{|J'_{12}|^2}{\Delta+U_{12}}-\frac{|J'_{12}|^2}{\Delta}$.
For  (a) $V_0'=6.2~E_r$,  (b) $6.0~E_r,$  and (c) 5.5~$E_r$, the periods of the $e^{-i\tau\hat{\cal H}_I/\hbar}$-induced oscillations are 3~ms, 5~ms, and 11~ms, respectively. As $V_0'$ decreases, these periods become longer.
On the other hand, the short period oscillations are attributed to the extra operation $\hat{\cal U}_{\rm ext}$ resulting from the breakdown of the perturbation.
This $\hat{\cal U}_{\rm ext}$ causes the resonant excitation of the intermediate state [shown in the middle panel in Fig.\,1\,(a) in the main text].
We can confirm this point from the oscillatory behavior of the two quantities $D$ and$N_{\rm 2nd}$.
For $V_0'=6.2~E_r, 6.0~E_r,$ and 5.5~$E_r$, the oscillation periods caused by $\hat{\cal U}_{\rm ext}$ are 0.5~ms, 0.35~ms, 0.17~ms, respectively.
As $V_0'$ decreases, the periods become shorter in contrast to those caused by  $e^{-i\tau\hat{\cal H}_I/\hbar}$.
In addition, we find that the amplitude of the $\hat{\cal U}_{\rm ext}$-induced oscillation become smaller.

\begin{figure}[t]
\includegraphics[width=9cm]{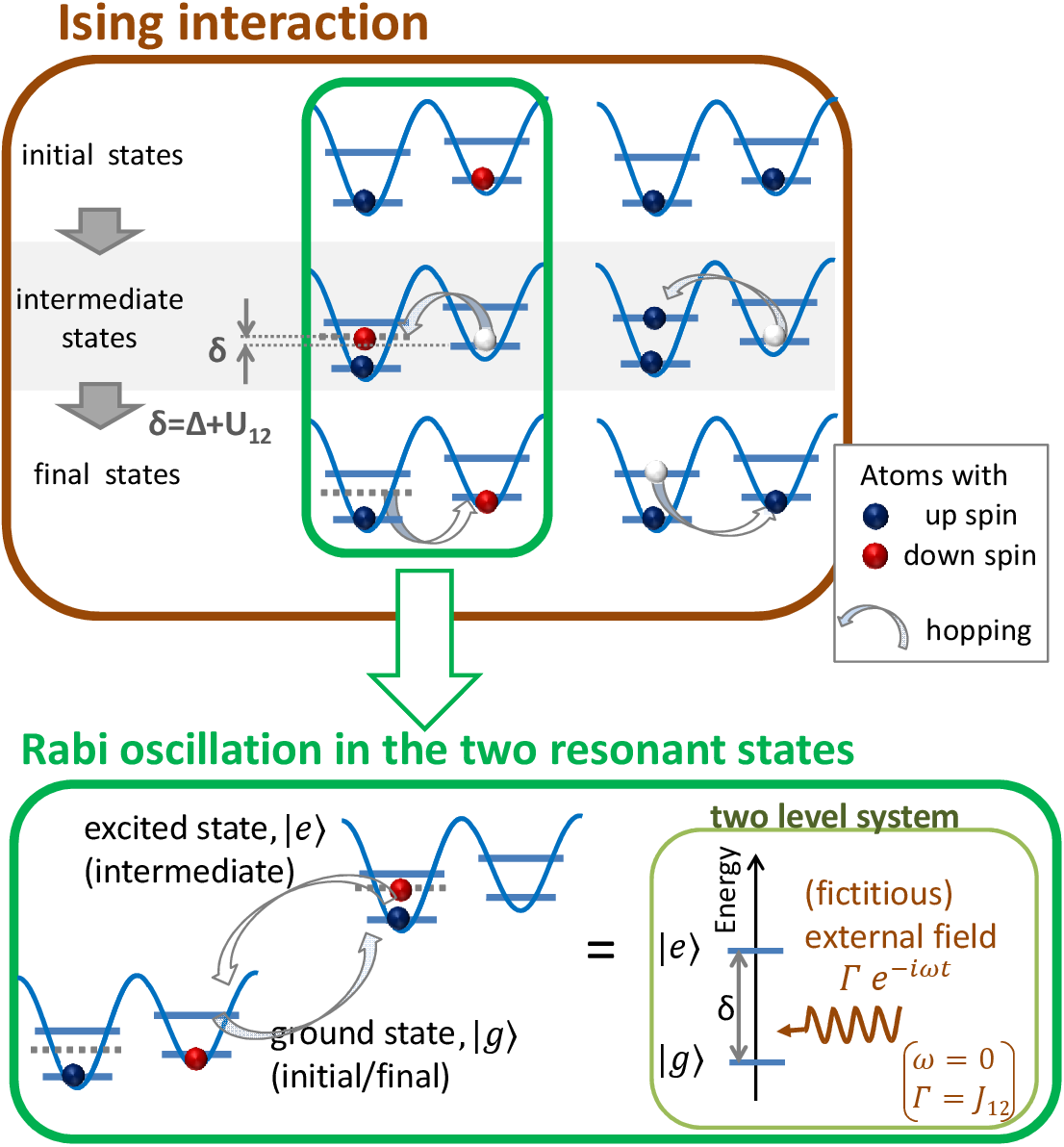}
\caption{ Schematic diagram representing the mechanisms of the Ising interaction discussed in the main text ({\it top panel}) and the Rabi-like oscillations ({\it bottom panel}).
The Ising interaction is resonantly enhanced by setting $\delta\equiv \Delta+U_{12}\sim0$.
The effective Rabi oscillation can be understood from the following points:
the two resonant states, $|g\rangle$ (the initial state) and $|e\rangle$ (the intermediate state), form the effective two level system; these two levels are coupled by the (fictitious) external field $\Gamma e^{-i\tau\omega}$ with $\omega=0$ and $\Gamma=J_{12}$; here, the energy difference $\delta$ corresponds to the detuning of the Rabi excitation.
}
\label{fig_schrabi}
\end{figure}

The oscillatory behavior caused by $\hat{\cal U}_{\rm ext}$ can be understood from the Rabi-oscillation-like mechanism as follows.
Figure~\ref{fig_schrabi} schematically shows the Rabi oscillation process that is included in the perturbative processes inducing the Ising interaction.
Here, the Ising interaction is resonantly enhanced; and as a result, the two states, the initial state and the intermediate state, have energies nearly degenerate with each other, where the energy difference is given by $\delta(=\Delta+U_{12})$. %, and $\delta=0$ corresponds to the resonant condition.
The time evolutions of such resonant states are characterized by the well-known Rabi oscillation.
As shown in Fig.~\ref{fig_schrabi},
two resonant states form the effective two level system (TLS) described by the 2$\times$2 Hamiltonian $\hat{\cal H}_{\rm TLS}=\left(\begin{array}{cc}0&J_{21}\\J_{12}&\delta\end{array}\right)$, and the eigenenergies are given by $(\delta\pm\sqrt{\delta^2+4|J_{12}|^2})/2$.
Thus, we can conclude that the time period of the oscillation caused by $\hat{\cal U}_{\rm ext}$ can be determined from the effective Rabi frequency given by $\hbar\Omega_R=\sqrt{\delta_{}^2+4|J_{12}|^{2}}$ with $\delta_{}$ corresponding to the detuning from the resonant condition $\delta_{}=0$.
In addition, the amplitude of the $\hat{\cal U}_{\rm ext}$-caused oscillation, which corresponds to the amplitudes of $D$ and $N_{\rm 2nd}$, is determined from $|J_{12}/\hbar\Omega_R|$.
Namely, the oscillations of $D$ and $N_{\rm 2nd}$ can be written as $\propto |J_{12}/\hbar\Omega_R|^2[1-\cos(\Omega_R \tau)]$.

We clarify that the above discussions can be confirmed from Fig.~\ref{fig_fid_rabi}.
As $V_0'$ varies from 6.2~$E_r$ (a) to 5.5~$E_r$ (c), the detuning $\delta$ changes from nearly zero to a larger value.
On the other hand, the interorbital hopping $J_{12}$ increases, while the change in $J_{12}$ is much smaller than that in $\delta$.
This means that, as $V_0'$ decreases, the Ising interaction $J_I$ decreases, while the Rabi frequency $\hbar\Omega_R$ increases.
Therefore, from the panel (a) to (c) in Fig.~\ref{fig_fid_rabi}, we find that the periods of $e^{-i\tau\hat{\cal H}_I/\hbar}$ and $\hat{\cal U}_{\rm ext}$-caused oscillations become longer and shorter, respectively, and the amplitude of the $\hat{\cal U}_{\rm ext}$-caused oscillation and those of $D$ and $N_{\rm 2nd}$ decrease.

Importantly, if the perturbative processes include the spin exchange processes, such a simple way to describe $\hat{\cal U}_{\rm ext}$-induced oscillation does not work well.
As discussed in the main text, a simple extra operation $\hat{\cal U}_{\rm ext}$ causes only one type of excitations, and this simple oscillatory dynamics can be controlled.
In contrast, when $\hat{\cal U}_{\rm ext}$ further includes $\hat{\cal H}_{ex}$, complex oscillatory dynamics will appear in the $F$ curves, causing great difficulty for the phase tuning.
Thus, we should note that the key to our method for overcoming the problem of the breakdown of the perturbation is to naturally create the Ising interaction.

\section{Phase-tuning scheme}\label{appD}
\begin{figure}[h]
\includegraphics[width=9cm]{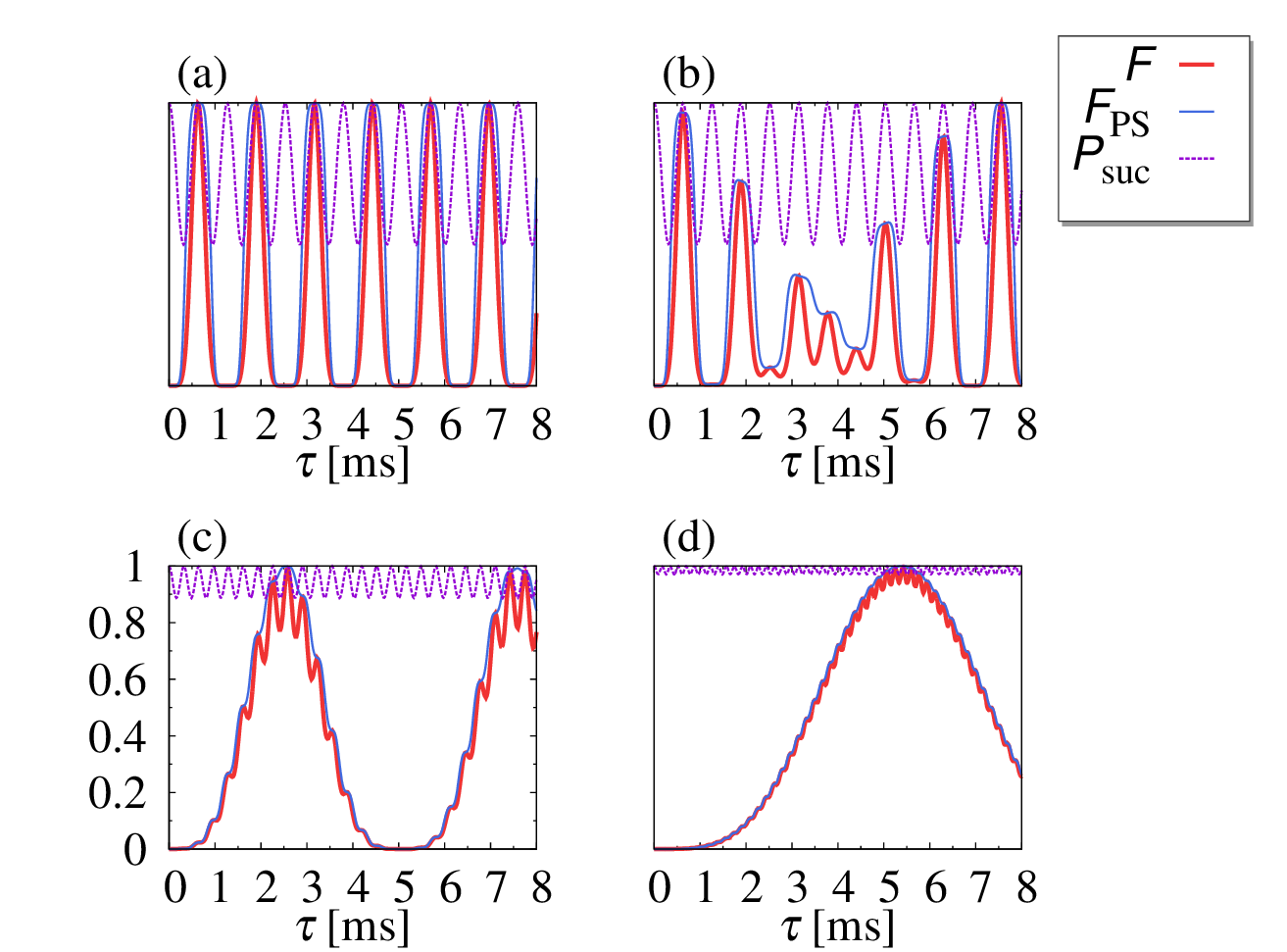}
\caption{
Simulation results for the 2-qubit system with $V_0=10\,E_r$ for $V_0'=6.48\,E_r$, 6.5\,$E_r$, 6.0\,$E_r$, and 5.5\,$E_r$.
}
\label{fig_rabi_PS}
\end{figure}

We discuss the details of the following phase-tuning scheme. % of the oscillatory behavior of the fidelity $F$ curves.
By controlling two oscillations caused by $\hat{\cal U}_{\rm ext}$ and $e^{-i\tau \hat{\cal H}_I/\hbar}$, we can minimize a decrease in fidelity, which occurs in return for shortening the operation time.
As mentioned in the main text, fidelity $F$ drastically changes as periods of the two oscillations vary; and when the two oscillations become inphase, the fidelity reaches a maximum value.
As discussed above in Sec.~C, these periods can be determined from the inverse of $J_I$ and $\hbar\Omega_R$.
Hence, the maximization of $F$ can be achieved under the following phase-tuning condition:
\begin{equation}
\ell J_I=(m-1/2)\hbar\Omega_R, \label{eq_ptc}
\end{equation}
where $\ell$ and $m$ are integers of $\ge 1$, and a factor of $1/2$ means a half cycle relative shift between the two oscillations.
Here, the operation time at which $F$ is maximum can be given by $(m-1/2)\pi\hbar/2J_I[\equiv (2m-1)\tau_I]$.
This phase tuning scheme can be easily implemented by tuning parameters such as $V_0'$.

In Fig.~\ref{fig_rabi_PS}, we show the fidelity $F$, the fidelity with the post selection $F_{\rm PS}$, and the success probability $P_{\rm suc}$ as a function of $\tau$ with varying $V_0'=6.48\,E_r$, 6.5\,$E_r$, 6.0\,$E_r$, and 5.5\,$E_r$.
The other parameters is set the same as those in Figs. 2 and 3 in the main text; $a=413$ nm, $a_{S}=-50$ nm, and $V_0=10 E_r$.
The above simulation results satisfy the phase-tuning condition with (a) $\ell=1$ and $m=1$, (b) $\ell=12$ and $m=7$, (c) $\ell=8$ and $m=1$, and (d) $\ell=31$ and $m=1$.
Note that, under all of the phase-tuning conditions shown here, we can achieve very high fidelity $F>0.99$ and $F_{\rm PS}\sim1$ with $P_{\rm suc}>0.99$.
By comparing Fig.~\ref{fig_rabi_PS}~(a), (b), and (d) [see also Fig.~3~(a) in the main text], we can conclude that, thanks to the phase tuning, the fidelity does not decrease even though we shorten the operation time from 5.5~ms to 0.6~ms.

Let us discuss the results in Fig.~\ref{fig_rabi_PS}~(a) and (b) in more detail.
For the simulation in Fig.~\ref{fig_rabi_PS}~ (a), we use the phase-tuning condition with $\ell=1$ and $m=1$ ($2J_I=\hbar\Omega_R$), giving the shortest operation time $\tau_I \sim 0.6$\,ms.
At first glance, we find that the $F$ curve in Fig.~\ref{fig_rabi_PS}~(a) has a single oscillation caused by $e^{-i\tau\hat{\cal H}_I/\hbar}$; however, the second oscillation resulting from $\hat{\cal U}_{\rm ext}$ is still alive, which can be clearly seen in the oscillation of $P_{\rm suc}$. %the condition $\ell=1$ and $m=1$ makes the oscillation
Achieving this shortest-time phase-tuning condition is accompanied by the appearance of the extra Rabi-like oscillations with a large amplitude; as a result, the oscillatory behavior of $P_{\rm suc}$ has a large amplitude.
Figure~\ref{fig_rabi_PS}~(b) shows the simulation results in which we slightly change $V_0'$ from that used in Fig.~\ref{fig_rabi_PS}~ (a).
This change in $V_0'$ violates the above condition $2J_I=\hbar\Omega_R$, leading to a decrease in fidelity at $\tau\sim 0.6$~ms; as a result, we obtain $F=0.97$ or $F_{\rm PS}\sim 0.97$ with $P_{\rm suc}\sim 0.99$ at $\tau\sim 0.6$~ms.
On the other hand, another condition with $\ell=12$ and $m=7$ is satisfied ($2J_I=13/12\hbar\Omega_R$), and thus we obtain very high fidelity $F>0.99$ at $\tau \sim 13\times0.6=7.8$.
Figure~\ref{fig_rabi_PS}~(a) and (b) highlights the importance of the phase tuning scheme to achieve high fidelity in a short operation time.

Next, we discuss the other simulations shown in Fig.~\ref{fig_rabi_PS}~(c) and (d), where we use large $\ell= 8$ and $31$ with $m=1$, respectively.
As $\ell$ increases with a fixed $m$, $J_I$ becomes much smaller than $\hbar\Omega_R$ ($J_I\ll \hbar\Omega_R$).
Thus, the operation time becomes longer, while the amplitude of the oscillation caused by $\hat{\cal U}_{\rm ext}$ becomes smaller.
However, as shown in  Fig.~\ref{fig_rabi_PS}~(d), a decrease in fidelity $F$ caused by $\hat{\cal U}_{\rm ext}$ is still beyond 5\%, indicating that the phase-tuning scheme is still required for the measurement based quantum computation (MBQC), even though we need to take an operation time $5.5$\,ms that is about ten times longer than the shortest one of 0.6\,ms.
Note that, since the periods of $\hat{\cal U}_{\rm ext}$-caused oscillation is fast, the accurate phase-tuning is required.
On the other hand, the single application of the postselection scheme without the phase tuning is effective when $J_I \ll \hbar\Omega_R$, because $F_{\rm PS}$ smoothly changes and reaches $\sim1$, which is, however, accompanied by a decrease in the success probability $P_{\rm suc}$ down to 0.95.

\subsubsection*{Long-time operation without phase tuning }
\begin{figure}[h]
\includegraphics[width=9cm]{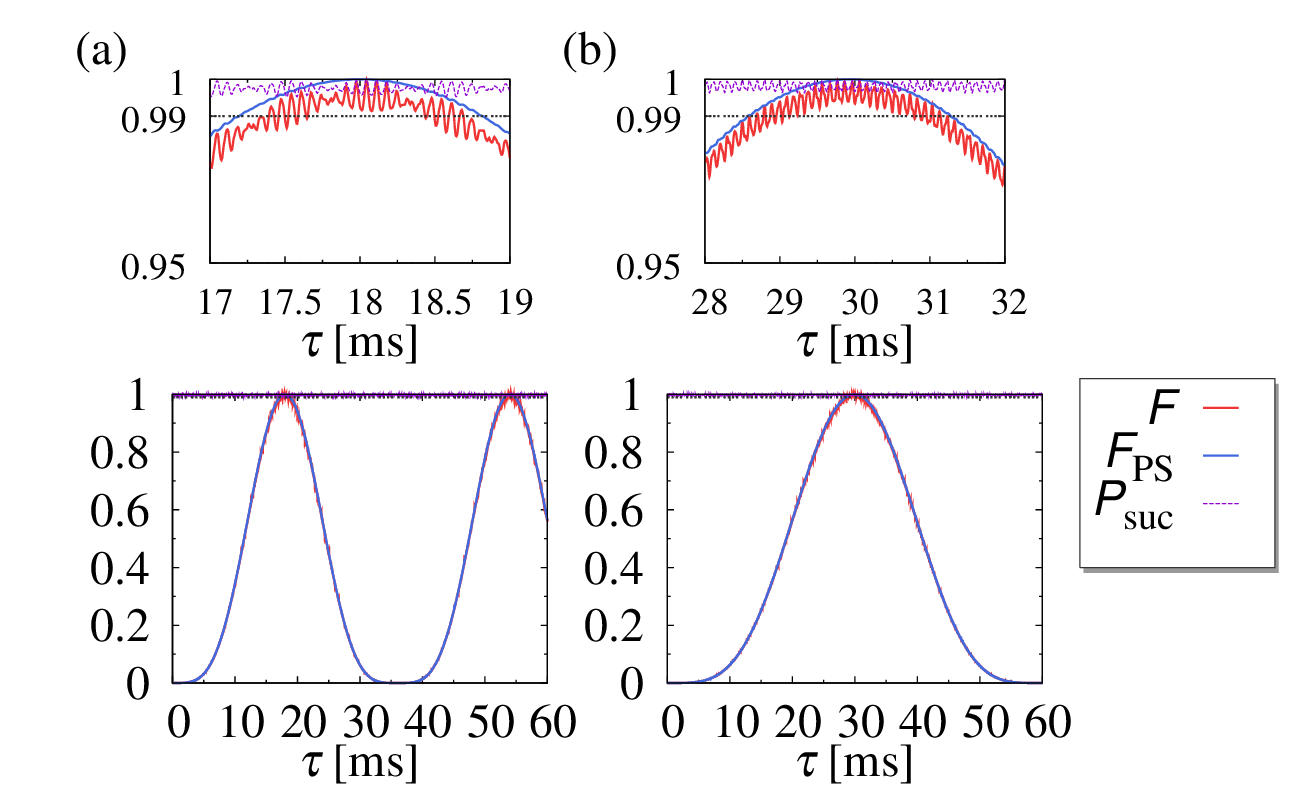}
\caption{
Simulation results for the 2-qubit system with $V_0=10\,E_r$ for $V_0'=4.0\,E_r$, and 3.0\,$E_r$.
The upper panels represent enlarged figures at around the maximum fidelity.
}
\label{fig_ideal}
\end{figure}

Let us further discuss the limit, in which the phase-tuning and postselection schemes are not required for achieving high fidelity, and surely the operation time becomes longer.
Namely, we examine the threshold limit achieving $F\gtrsim0.99$ without a phase tuning.
As mentioned in the above, the extra Rabi oscillations caused by $\hat{\cal U}_{\rm ext}$ can be suppressed by getting away from the resonant condition $\delta_{}\equiv \Delta+U_{12} \sim 0$.
Then, finally, we can achieve the ideal perturbation limit without any extra Rabi oscillations, where $\hat{\cal U}_{\rm ext}$ is equivalent to the identity operator~$\hat{1}$.
Figure \ref{fig_ideal} shows the simulation results with $V_0'=$4.0\,$E_r$ and 3.0~$E_r$.
The panel (a) shows a marginal of the threshold limit, while the panel (b) achieves enough the limit.
Here, two fidelity curves $F$ and $F_{\rm PS}$ overlap with each other, and $P_{\rm suc}$ almost equals to 1 for any $\tau$.
The operation times are taken about 18~ms and 30~ms, respectively.

We should note that this discussion cannot be naively applied to the multipartite $n$-qubit entanglement generations with $n>2$.
This is because, for $n$-qubit systems with $n>2$, the corrective excitations introduce another types of $\hat{\cal U}_{\rm ext}$ as mentioned in the main text.
On the other hand, by comparing these results with the results shown in the previous section, we can see how does the phase-tuning scheme allow us to shorten the operation time without the time-fidelity tradeoff resulting from the breakdown of the perturbation: The operation time can be shortened from 30~ms to 0.6~ms.

 %, which caused by the enhancement of the Ising interaction.

\section{Duration time of being high fidelity}\label{appE}

This section provides a discussion about the duration time in which we can keep the fidelity of cluster states beyond the threshold $\gtrsim 0.99$.
The estimated duration time naively gives an indication of the experimental time sequence;
Within this duration time, the additional two-site period potential will be turned off ($V_0'\to0$).

\begin{figure*}
\includegraphics[width=14cm]{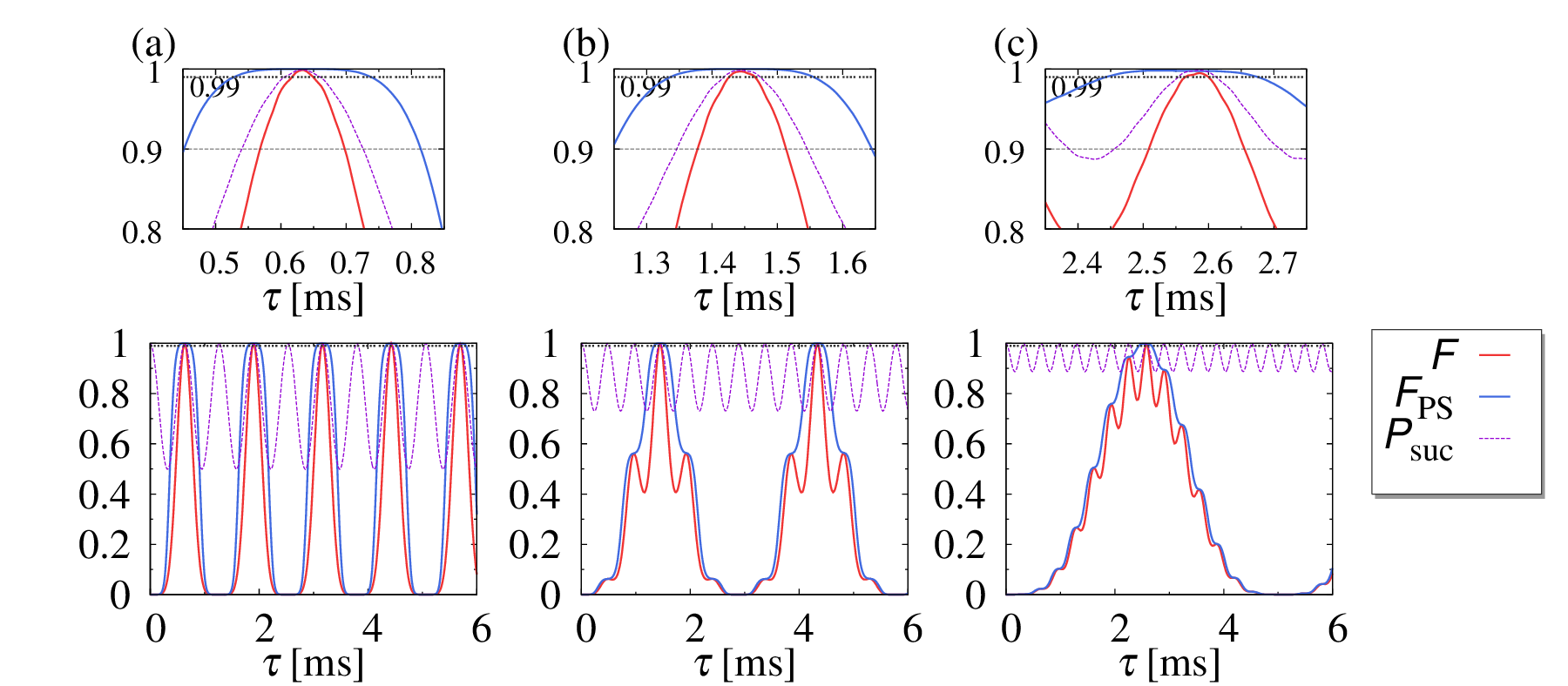}
\caption{
Simulation results for the 2-qubit system with $V_0=10\,E_r$ for $V_0'=6.48\,E_r$, 6.2\,$E_r$, and 6.0\,$E_r$.
The upper panels represent enlarged figures at around the maximum fidelity.
}
\label{fig_duration}
\end{figure*}

In the bottom panels in Fig.~\ref{fig_duration}, we show the same simulation results in Fig.~\ref{fig_rabi_PS} and also in Fig.~3 in the main text, while in the top panels, we present the enlarged figures at around $\tau_I$.
We show the results for various phase-tuning conditions for $\ell=1, 3$, and 8 with $m=1$, corresponding to the various $V_0'=6.48\,E_r$, 6.2\,$E_r$, and 6.0\,$E_r$, respectively.
A very high fidelity beyond the threshold $F\gtrsim0.99$ can be achieved in the duration of about 0.04~ms for all $V_0'$.
Note that the $\ell$ dependence of the duration time is small but finite, and the duration time becomes longer as $V_0'$ decreases.
We can find that the postselection scheme allows us to enlarge the duration time.
The duration of being $F_{\rm PS}\gtrsim 0.99$ is about 0.21~ms, 0.22~ms, and 0.23~ms for $V_0'=6.48\,E_r$, 6.2\,$E_r$, and 6.0\,$E_r$, respectively.
We can expect that this advantage of the post selection scheme will make a practical progress in experiments.
On the other hand, in return for lengthening duration time, the success probability $P_{\rm suc}$ decreases down to 0.85, 0.87, and 0.9, respectively.
Here, we should note that our schemes are suitable for the loss tolerant MBQC scheme, because MBQC is more robust against losses than errors \cite{Stace2009,Barrett2010}.
The failure events of the measurement used in the postselection scheme can be regarded as losses of qubits.
Therefore, it is practically useful to enhance the fidelity and enlarge duration time in return for the increasing losses.

  %1.26 2.89 5.05

\begin{thebibliography}{26}
\expandafter\ifx\csname natexlab\endcsname\relax\def\natexlab#1{#1}\fi
\expandafter\ifx\csname bibnamefont\endcsname\relax
  \def\bibnamefont#1{#1}\fi
\expandafter\ifx\csname bibfnamefont\endcsname\relax
  \def\bibfnamefont#1{#1}\fi
\expandafter\ifx\csname citenamefont\endcsname\relax
  \def\citenamefont#1{#1}\fi
\expandafter\ifx\csname url\endcsname\relax
  \def\url#1{\texttt{#1}}\fi
\expandafter\ifx\csname urlprefix\endcsname\relax\def\urlprefix{URL }\fi
\providecommand{\bibinfo}[2]{#2}
\providecommand{\eprint}[2][]{\url{#2}}

\bibitem[{\citenamefont{Briegel et~al.}(2009)\citenamefont{Briegel, Browne,
  Dur, Raussendorf, and Van~den Nest}}]{Briegel2009}
\bibinfo{author}{\bibfnamefont{H.~J.} \bibnamefont{Briegel}},
  \bibinfo{author}{\bibfnamefont{D.~E.} \bibnamefont{Browne}},
  \bibinfo{author}{\bibfnamefont{W.}~\bibnamefont{Dur}},
  \bibinfo{author}{\bibfnamefont{R.}~\bibnamefont{Raussendorf}},
  \bibnamefont{and} \bibinfo{author}{\bibfnamefont{M.}~\bibnamefont{Van~den
  Nest}}, \bibinfo{journal}{Nat. Phys.} \textbf{\bibinfo{volume}{5}},
  \bibinfo{pages}{19} (\bibinfo{year}{2009}).

\bibitem[{\citenamefont{Raussendorf and Briegel}(2001)}]{Raussendorf2001}
\bibinfo{author}{\bibfnamefont{R.}~\bibnamefont{Raussendorf}} \bibnamefont{and}
  \bibinfo{author}{\bibfnamefont{H.~J.} \bibnamefont{Briegel}},
  \bibinfo{journal}{Phys. Rev. Lett.} \textbf{\bibinfo{volume}{86}},
  \bibinfo{pages}{5188} (\bibinfo{year}{2001}).

\bibitem[{\citenamefont{Raussendorf et~al.}(2007)\citenamefont{Raussendorf,
  Harrington, and Goyal}}]{Raussendorf2007}
\bibinfo{author}{\bibfnamefont{R.}~\bibnamefont{Raussendorf}},
  \bibinfo{author}{\bibfnamefont{J.}~\bibnamefont{Harrington}},
  \bibnamefont{and} \bibinfo{author}{\bibfnamefont{K.}~\bibnamefont{Goyal}},
  \bibinfo{journal}{New J. Phys.} \textbf{\bibinfo{volume}{9}},
  \bibinfo{pages}{199} (\bibinfo{year}{2007}).

\bibitem[{\citenamefont{Raussendorf and Harrington}(2007)}]{Raussendorf2007a}
\bibinfo{author}{\bibfnamefont{R.}~\bibnamefont{Raussendorf}} \bibnamefont{and}
  \bibinfo{author}{\bibfnamefont{J.}~\bibnamefont{Harrington}},
  \bibinfo{journal}{Phys. Rev. Lett.} \textbf{\bibinfo{volume}{98}},
  \bibinfo{pages}{190504} (\bibinfo{year}{2007}).

\bibitem[{\citenamefont{Bloch}(2008)}]{Bloch2008a}
\bibinfo{author}{\bibfnamefont{I.}~\bibnamefont{Bloch}},
  \bibinfo{journal}{Nature (London)} \textbf{\bibinfo{volume}{453}},
  \bibinfo{pages}{1016} (\bibinfo{year}{2008}).

\bibitem[{\citenamefont{Bakr et~al.}(2010)\citenamefont{Bakr, Peng, Tai, Ma,
  Simon, Gillen, F\"olling, Pollet, and Greiner}}]{Bakr2010}
\bibinfo{author}{\bibfnamefont{W.~S.} \bibnamefont{Bakr}},
  \bibinfo{author}{\bibfnamefont{A.}~\bibnamefont{Peng}},
  \bibinfo{author}{\bibfnamefont{M.~E.} \bibnamefont{Tai}},
  \bibinfo{author}{\bibfnamefont{R.}~\bibnamefont{Ma}},
  \bibinfo{author}{\bibfnamefont{J.}~\bibnamefont{Simon}},
  \bibinfo{author}{\bibfnamefont{J.~I.} \bibnamefont{Gillen}},
  \bibinfo{author}{\bibfnamefont{S.}~\bibnamefont{F\"olling}},
  \bibinfo{author}{\bibfnamefont{L.}~\bibnamefont{Pollet}}, \bibnamefont{and}
  \bibinfo{author}{\bibfnamefont{M.}~\bibnamefont{Greiner}},
  \bibinfo{journal}{Science} \textbf{\bibinfo{volume}{329}},
  \bibinfo{pages}{547} (\bibinfo{year}{2010}).

\bibitem[{\citenamefont{Weitenberg et~al.}(2011)\citenamefont{Weitenberg,
  Endres, Sherson, Cheneau, Schausz, Fukuhara, Bloch, and
  Kuhr}}]{Weitenberg2011}
\bibinfo{author}{\bibfnamefont{C.}~\bibnamefont{Weitenberg}},
  \bibinfo{author}{\bibfnamefont{M.}~\bibnamefont{Endres}},
  \bibinfo{author}{\bibfnamefont{J.~F.} \bibnamefont{Sherson}},
  \bibinfo{author}{\bibfnamefont{M.}~\bibnamefont{Cheneau}},
  \bibinfo{author}{\bibfnamefont{P.}~\bibnamefont{Schausz}},
  \bibinfo{author}{\bibfnamefont{T.}~\bibnamefont{Fukuhara}},
  \bibinfo{author}{\bibfnamefont{I.}~\bibnamefont{Bloch}}, \bibnamefont{and}
  \bibinfo{author}{\bibfnamefont{S.}~\bibnamefont{Kuhr}},
  \bibinfo{journal}{Nature (London)} \textbf{\bibinfo{volume}{471}},
  \bibinfo{pages}{319} (\bibinfo{year}{2011}).

\bibitem[{\citenamefont{Mandel et~al.}(2003{\natexlab{a}})\citenamefont{Mandel,
  Greiner, Widera, Rom, Hansch, and Bloch}}]{Mandel2003a}
\bibinfo{author}{\bibfnamefont{O.}~\bibnamefont{Mandel}},
  \bibinfo{author}{\bibfnamefont{M.}~\bibnamefont{Greiner}},
  \bibinfo{author}{\bibfnamefont{A.}~\bibnamefont{Widera}},
  \bibinfo{author}{\bibfnamefont{T.}~\bibnamefont{Rom}},
  \bibinfo{author}{\bibfnamefont{T.~W.} \bibnamefont{Hansch}},
  \bibnamefont{and} \bibinfo{author}{\bibfnamefont{I.}~\bibnamefont{Bloch}},
  \bibinfo{journal}{Phys. Rev. Lett.} \textbf{\bibinfo{volume}{91}},
  \bibinfo{pages}{010407} (\bibinfo{year}{2003}{\natexlab{a}}).

\bibitem[{\citenamefont{Mandel et~al.}(2003{\natexlab{b}})\citenamefont{Mandel,
  Greiner, Widera, Rom, Hansch, and Bloch}}]{Mandel2003}
\bibinfo{author}{\bibfnamefont{O.}~\bibnamefont{Mandel}},
  \bibinfo{author}{\bibfnamefont{M.}~\bibnamefont{Greiner}},
  \bibinfo{author}{\bibfnamefont{A.}~\bibnamefont{Widera}},
  \bibinfo{author}{\bibfnamefont{T.}~\bibnamefont{Rom}},
  \bibinfo{author}{\bibfnamefont{T.~W.} \bibnamefont{Hansch}},
  \bibnamefont{and} \bibinfo{author}{\bibfnamefont{I.}~\bibnamefont{Bloch}},
  \bibinfo{journal}{Nature (London)} \textbf{\bibinfo{volume}{425}},
  \bibinfo{pages}{937} (\bibinfo{year}{2003}{\natexlab{b}}).

\bibitem[{\citenamefont{Trotzky et~al.}(2008)\citenamefont{Trotzky, Cheinet,
  Folling, Feld, Schnorrberger, Rey, Polkovnikov, Demler, Lukin, and
  Bloch}}]{Trotzky2008}
\bibinfo{author}{\bibfnamefont{S.}~\bibnamefont{Trotzky}},
  \bibinfo{author}{\bibfnamefont{P.}~\bibnamefont{Cheinet}},
  \bibinfo{author}{\bibfnamefont{S.}~\bibnamefont{Folling}},
  \bibinfo{author}{\bibfnamefont{M.}~\bibnamefont{Feld}},
  \bibinfo{author}{\bibfnamefont{U.}~\bibnamefont{Schnorrberger}},
  \bibinfo{author}{\bibfnamefont{A.~M.} \bibnamefont{Rey}},
  \bibinfo{author}{\bibfnamefont{A.}~\bibnamefont{Polkovnikov}},
  \bibinfo{author}{\bibfnamefont{E.~A.} \bibnamefont{Demler}},
  \bibinfo{author}{\bibfnamefont{M.~D.} \bibnamefont{Lukin}}, \bibnamefont{and}
  \bibinfo{author}{\bibfnamefont{I.}~\bibnamefont{Bloch}},
  \bibinfo{journal}{Science} \textbf{\bibinfo{volume}{319}},
  \bibinfo{pages}{295} (\bibinfo{year}{2008}).

\bibitem[{\citenamefont{Trotzky et~al.}(2010)\citenamefont{Trotzky, Chen,
  Schnorrberger, Cheinet, and Bloch}}]{Trotzky2010}
\bibinfo{author}{\bibfnamefont{S.}~\bibnamefont{Trotzky}},
  \bibinfo{author}{\bibfnamefont{Y.-A.} \bibnamefont{Chen}},
  \bibinfo{author}{\bibfnamefont{U.}~\bibnamefont{Schnorrberger}},
  \bibinfo{author}{\bibfnamefont{P.}~\bibnamefont{Cheinet}}, \bibnamefont{and}
  \bibinfo{author}{\bibfnamefont{I.}~\bibnamefont{Bloch}},
  \bibinfo{journal}{Phys. Rev. Lett.} \textbf{\bibinfo{volume}{105}},
  \bibinfo{pages}{265303} (\bibinfo{year}{2010}).

\bibitem{Anderlini2007CEI}
M. Anderlini, P.~J. Lee, B.~L. Brown, J. Sebby-Strabley, W.~D. Phillips and J.~V. Porto, Nature (London) {\bf 448}, 452-456 (2007).


\bibitem[{\citenamefont{Vaucher et~al.}(2008)\citenamefont{Vaucher, Nunnenkamp,
  and Jaksch}}]{Vaucher2008}
\bibinfo{author}{\bibfnamefont{B.}~\bibnamefont{Vaucher}},
  \bibinfo{author}{\bibfnamefont{A.}~\bibnamefont{Nunnenkamp}},
  \bibnamefont{and} \bibinfo{author}{\bibfnamefont{D.}~\bibnamefont{Jaksch}},
  \bibinfo{journal}{New J. Phys.} \textbf{\bibinfo{volume}{10}},
  \bibinfo{pages}{023005} (\bibinfo{year}{2008}).

\bibitem[{\citenamefont{Duan et~al.}(2003)\citenamefont{Duan, Demler, and
  Lukin}}]{Duan2003}
\bibinfo{author}{\bibfnamefont{L.-M.} \bibnamefont{Duan}},
  \bibinfo{author}{\bibfnamefont{E.}~\bibnamefont{Demler}}, \bibnamefont{and}
  \bibinfo{author}{\bibfnamefont{M.~D.} \bibnamefont{Lukin}},
  \bibinfo{journal}{Phys. Rev. Lett.} \textbf{\bibinfo{volume}{91}},
  \bibinfo{pages}{090402} (\bibinfo{year}{2003}).

\bibitem[{\citenamefont{Jiang et~al.}(2009)\citenamefont{Jiang, Rey,
  Romero-Isart, Garcia-Ripoll, Sanpera, and Lukin}}]{Jiang2009}
\bibinfo{author}{\bibfnamefont{L.}~\bibnamefont{Jiang}},
  \bibinfo{author}{\bibfnamefont{A.~M.} \bibnamefont{Rey}},
  \bibinfo{author}{\bibfnamefont{O.}~\bibnamefont{Romero-Isart}},
  \bibinfo{author}{\bibfnamefont{J.~J.} \bibnamefont{Garcia-Ripoll}},
  \bibinfo{author}{\bibfnamefont{A.}~\bibnamefont{Sanpera}}, \bibnamefont{and}
  \bibinfo{author}{\bibfnamefont{M.~D.} \bibnamefont{Lukin}},
  \bibinfo{journal}{Phys. Rev. A} \textbf{\bibinfo{volume}{79}},
  \bibinfo{pages}{022309} (\bibinfo{year}{2009}).

\bibitem[{\citenamefont{Chen et~al.}(2011)\citenamefont{Chen, Nascimb\`ene,
  Aidelsburger, Atala, Trotzky, and Bloch}}]{YuAo2011}
\bibinfo{author}{\bibfnamefont{Y.-A.} \bibnamefont{Chen}},
  \bibinfo{author}{\bibfnamefont{S.}~\bibnamefont{Nascimb\`ene}},
  \bibinfo{author}{\bibfnamefont{M.}~\bibnamefont{Aidelsburger}},
  \bibinfo{author}{\bibfnamefont{M.}~\bibnamefont{Atala}},
  \bibinfo{author}{\bibfnamefont{S.}~\bibnamefont{Trotzky}}, \bibnamefont{and}
  \bibinfo{author}{\bibfnamefont{I.}~\bibnamefont{Bloch}},
  \bibinfo{journal}{Phys. Rev. Lett.} \textbf{\bibinfo{volume}{107}},
  \bibinfo{pages}{210405} (\bibinfo{year}{2011}).

\bibitem{Simon2011QSO}
J. Simon, W.~S. Bakr, R. Ma, M.~E. Tai, P.~M. Preiss and M. Greiner, Nature (London) {\bf 472}, 307-312 (2011).



\bibitem{Anderson}
P. W. Anderson, Phys. Rev. {\bf 115}, 2 (1959).





\bibitem[{\citenamefont{Regal and Jin}(2003)}]{Regal2003}
\bibinfo{author}{\bibfnamefont{C.~A.} \bibnamefont{Regal}} \bibnamefont{and}
  \bibinfo{author}{\bibfnamefont{D.~S.} \bibnamefont{Jin}},
  \bibinfo{journal}{Phys. Rev. Lett.} \textbf{\bibinfo{volume}{90}},
  \bibinfo{pages}{230404} (\bibinfo{year}{2003}).



\bibitem[{Not({\natexlab{b}})}]{NoteBell}
\bibinfo{note}{In the two-qubit system, the cluster state $|{\rm CS}\rangle$ is
  equivalent to the Bell pair.}





\bibitem[{\citenamefont{Rom et~al.}(2004)\citenamefont{Rom, Best, Mandel,
  Widera, Greiner, H\"ansch, and Bloch}}]{Rom2004}
\bibinfo{author}{\bibfnamefont{T.}~\bibnamefont{Rom}},
  \bibinfo{author}{\bibfnamefont{T.}~\bibnamefont{Best}},
  \bibinfo{author}{\bibfnamefont{O.}~\bibnamefont{Mandel}},
  \bibinfo{author}{\bibfnamefont{A.}~\bibnamefont{Widera}},
  \bibinfo{author}{\bibfnamefont{M.}~\bibnamefont{Greiner}},
  \bibinfo{author}{\bibfnamefont{T.~W.} \bibnamefont{H\"ansch}},
  \bibnamefont{and} \bibinfo{author}{\bibfnamefont{I.}~\bibnamefont{Bloch}},
  \bibinfo{journal}{Phys. Rev. Lett.} \textbf{\bibinfo{volume}{93}},
  \bibinfo{pages}{073002} (\bibinfo{year}{2004}).

\bibitem[{\citenamefont{Sugawa et~al.}(2011)\citenamefont{Sugawa, Inaba, Taie,
  Yamazaki, Yamashita, and Takahashi}}]{Sugawa2011}
\bibinfo{author}{\bibfnamefont{S.}~\bibnamefont{Sugawa}},
  \bibinfo{author}{\bibfnamefont{K.}~\bibnamefont{Inaba}},
  \bibinfo{author}{\bibfnamefont{S.}~\bibnamefont{Taie}},
  \bibinfo{author}{\bibfnamefont{R.}~\bibnamefont{Yamazaki}},
  \bibinfo{author}{\bibfnamefont{M.}~\bibnamefont{Yamashita}},
  \bibnamefont{and}
  \bibinfo{author}{\bibfnamefont{Y.}~\bibnamefont{Takahashi}},
  \bibinfo{journal}{Nat. Phys.} \textbf{\bibinfo{volume}{7}},
  \bibinfo{pages}{642} (\bibinfo{year}{2011}).

\bibitem[{\citenamefont{Bakr et~al.}(2011)\citenamefont{Bakr, Preiss, Tai, Ma,
  Simon, and Greiner}}]{Bakr2011}
\bibinfo{author}{\bibfnamefont{W.~S.} \bibnamefont{Bakr}},
  \bibinfo{author}{\bibfnamefont{P.~M.} \bibnamefont{Preiss}},
  \bibinfo{author}{\bibfnamefont{M.~E.} \bibnamefont{Tai}},
  \bibinfo{author}{\bibfnamefont{R.}~\bibnamefont{Ma}},
  \bibinfo{author}{\bibfnamefont{J.}~\bibnamefont{Simon}}, \bibnamefont{and}
  \bibinfo{author}{\bibfnamefont{M.}~\bibnamefont{Greiner}},
  \bibinfo{journal}{Nature (London)} \textbf{\bibinfo{volume}{480}},
  \bibinfo{pages}{500} (\bibinfo{year}{2011}).



\bibitem[{Not({\natexlab{c}})}]{NoteGradient}
\bibinfo{note}{
The gradient potential $gx$ may excite atoms from the lowest orbital to the higher orbitals. Such an excitation can be suppressed in the deep lattice (large $V_0$) with a small gradient $g$.
Note that the gradient $g$ depends on $V_0'$ to satisfy the resonance condition.
We thus set $V_0 - V_0'$ to be large in the simulation.
}





\bibitem[{\citenamefont{Stace et~al.}(2009)\citenamefont{Stace, Barrett, and
  Doherty}}]{Stace2009}
\bibinfo{author}{\bibfnamefont{T.~M.} \bibnamefont{Stace}},
  \bibinfo{author}{\bibfnamefont{S.~D.} \bibnamefont{Barrett}},
  \bibnamefont{and} \bibinfo{author}{\bibfnamefont{A.~C.}
  \bibnamefont{Doherty}}, \bibinfo{journal}{Phys. Rev. Lett.}
  \textbf{\bibinfo{volume}{102}}, \bibinfo{pages}{200501}
  (\bibinfo{year}{2009}).

\bibitem[{\citenamefont{Barrett and Stace}(2010)}]{Barrett2010}
\bibinfo{author}{\bibfnamefont{S. D.}~\bibnamefont{Barrett}} \bibnamefont{and}
  \bibinfo{author}{\bibfnamefont{T. M.}~\bibnamefont{Stace}},
  \bibinfo{journal}{Phys. Rev. Lett.} \textbf{\bibinfo{volume}{105}},
  \bibinfo{pages}{200502} (\bibinfo{year}{2010}).







\bibitem[{\citenamefont{K\"{o}hl et~al.}(2005)\citenamefont{K\"{o}hl, Moritz,
  St\"{o}ferle, G\"{u}nter, and Esslinger}}]{Kohl2005}
\bibinfo{author}{\bibfnamefont{M.}~\bibnamefont{K\"{o}hl}},
  \bibinfo{author}{\bibfnamefont{H.}~\bibnamefont{Moritz}},
  \bibinfo{author}{\bibfnamefont{T.}~\bibnamefont{St\"{o}ferle}},
  \bibinfo{author}{\bibfnamefont{K.}~\bibnamefont{G\"{u}nter}},
  \bibnamefont{and}
  \bibinfo{author}{\bibfnamefont{T.}~\bibnamefont{Esslinger}},
  \bibinfo{journal}{Phys. Rev. Lett.} \textbf{\bibinfo{volume}{94}},
  \bibinfo{pages}{080403} (\bibinfo{year}{2005}).


\bibitem[{\citenamefont{Will et~al.}(2010)\citenamefont{Will, Best, Schneider,
  Hackermuller, Luhmann, and Bloch}}]{Will2010}
\bibinfo{author}{\bibfnamefont{S.}~\bibnamefont{Will}},
  \bibinfo{author}{\bibfnamefont{T.}~\bibnamefont{Best}},
  \bibinfo{author}{\bibfnamefont{U.}~\bibnamefont{Schneider}},
  \bibinfo{author}{\bibfnamefont{L.}~\bibnamefont{Hackermuller}},
  \bibinfo{author}{\bibfnamefont{D.-S.} \bibnamefont{Luhmann}},
  \bibnamefont{and} \bibinfo{author}{\bibfnamefont{I.}~\bibnamefont{Bloch}},
  \bibinfo{journal}{Nature (London)} \textbf{\bibinfo{volume}{465}},
  \bibinfo{pages}{197} (\bibinfo{year}{2010}).





\end{thebibliography}
\end{document}